\documentclass[12pt]{iopart}

\newcommand{\no}{\nonumber\\}
\newcommand{\be}{\begin{equation}}
\newcommand{\ee}{\end{equation}}
\newcommand{\ba}{\begin{eqnarray}}
\newcommand{\ea}{\end{eqnarray}}
\newcommand{\ci}[1]{\cite{#1}}
\newcommand{\la}[1]{\label{#1}}

\def\gl#1{(\ref{#1})}

\begin{document}
\title[Nonlinear Supersymmetry]{Nonlinear Supersymmetry for Spectral Design in Quantum Mechanics}
\author{
A A Andrianov\dag\ and F Cannata\ddag}
\address{\dag\ V A Fock Department of Theoretical Physics, Sankt-Petersburg  
State University,
198904 Sankt-Petersburg, Russia}
\address{\ddag\ 
Dipartmento di Fisica and INFN, Via Irnerio 46, 40126 Bologna,
Italy}

\ead{andrianov@bo.infn.it;\quad cannata@bo.infn.it}

%%%%%%%%%%%%%%%%%%%%%%%%%%%%%%%%%%%%%%%%%%%%%%%%%
\begin{abstract}

Nonlinear (Polynomial, N-fold) SUSY approach to preparation of quantum
systems with pre-planned spectral properties is reviewed. The full
classification of ladder-reducible and irreducible chains of SUSY
algebras in one-dimensional QM is given.  Possible extensions of SUSY
in one dimension are described. They  include  (no more than)  ${\cal
N} =2$ extended SUSY with two nilpotent SUSY charges which generate
the hidden symmetry acting as a  central charge. Embedding stationary
quantum systems into a non-stationary SUSY QM is shown to yield new
insight on quantum orbits and on spectrum generating algebras.

\end{abstract}

\submitto{Journal of Physics A: Math. Gen.}

\maketitle

\section{Introduction: Darboux intertwining, Schr\"odinger factorization,
Witten SUSY mechanics in one basket}

\hspace*{3ex} The concept of Supersymmetric Quantum Mechanics (SUSY QM) initially
associated to the $0 + 1$ dimensional SUSY Field Theory \cite{nico}  
aimed  at a simplified analysis of difficult 
problems of multi-dimensional QFT such as spontaneous 
SUSY breaking, vacuum properties beyond
perturbation theory etc. \cite{witt,salom}. In addition, the realization 
of different SUSY algebras in Quantum Physics turned out to be easily achieved
in certain QM models \cite{crom,akul}.

Soon after its formulation  SUSY QM was well identified with the Quantum
Mechanics of isospectral systems described by  Hamiltonians with
almost coincident energy spectra \cite{coop}--\cite{fern}.  
SUSY  manifested itself through
intertwining Darboux \cite{darb} 
transformations between isospectral partners, used before in 
Mathematics \cite{crum}--\cite{matv}. The latter property presumably gave
E.~Schr\"odinger \cite{schr} a hint to factorize the corresponding Hamiltonians
into a product of simplest, first-order Darboux operators \cite{inf}. 
Altogether the differential realization of SUSY in QM  
stimulated its application for the spectral design, 
{\it i.e.} for preparation of quantum potential systems
with 
pre-planned energy spectra \cite{abei}--\cite{kirch} and scattering data 
\cite{mnieto}, \cite{suku2}--\cite{samscat} or potential profiles 
\cite{suku3,kumar}
in a constructive way. For the spectral design-kit the non-linear supersymmetry
in its differential realization \cite{ais}--\cite{iof2004} 
has become one of the most efficient tools
to build various isospectral quantum systems with desired features.
By now, several books and reviews \cite{genkri}--\cite{spirid1} 
devoted to
certain achievements and diverse applications of
SUSY QM approach  illuminated at length many of 
the above mentioned 
trends. In contradistinction, the role and 
the structure of nonlinear
SUSY QM just developed in the last decade and has not been yet
surveyed.

The interplay between the algebraic and differential properties of
nonlinear supersymmetry, in the spectral design, is guiding 
our present work: we are going to
clarify the benefits of algebraic SUSY approach to the old Darboux-Crum method,
to develop, with these tools, the complete classification of differential 
realizations of non-linear SUSY algebras and to give insight to intrinsic
links between SUSY isospectrality and hidden symmetries in particular
quantum systems.

In what follows we restrict ourselves 
to the full analysis in the one-dimensional,
one-component, stationary QM elaborating few important 
examples for the 
non-stationary
Schr\"odinger equation and leave   
complex, matrix, multidimensional and/or relativistic
(Dirac, Klein-Gordon etc.) equations without any detailed comments.

Let us start the retrospection of  SUSY QM: 
consider two one-dimensional Schr\"odinger Hamiltonians $h^\pm$ defined on the 
line and assemble them into a matrix Super-Hamiltonian,
\be
H = \left(\begin{array}{cc}
h^+& 0\\
0 & h^-
\end{array}\right) =\left(\begin{array}{cc}
-\partial^2 + V_1(x)& 0\\
0 & - \partial^2 + V_2(x)
\end{array}\right), \qquad \partial \equiv d/dx , \label{hamil} 
\ee
with non-singular real potentials.
These Hamiltonians $h^\pm$ are supposed to have (almost) the same energy levels
for bound states and/or the same spectral densities for  continuum 
parts of spectra. Furthermore, assume that 
their  {\sl isospectral} connection  is provided by  intertwining
with the  Crum-Darboux \cite{darb,crum} transformation operators  $q^{\pm}_N$,
\be
h^+ q^+_N = q^+_N h^- , \quad q^-_N h^+ = h^- q^-_N , \label{intertw}
\ee
where  $N$-th order differential operators
\be
q^{\pm}_N =\sum_{k=0}^N w^{\pm}_k (x)\partial^k, \quad w^{\pm}_N = const
\equiv (\mp 1)^N. \label{crum}
\ee
The conventional, linear ${\cal N} = 1$ SUSY QM  in the fermion number
representation \cite{witt} is implemented by nilpotent supercharges
$Q_1,\ Q^{\dagger}_1$ of 
 first order in derivatives built from a real super-potential $\chi(x)$, 
\be
\fl
q^\pm_1 \equiv \mp\partial + \chi(x);\qquad \Longrightarrow \qquad 
Q_1=\left(\begin{array}{cc}
 0 &  q^+_1\\
0 & 0
\end{array}\right),\quad  Q^2_1 = \left(Q^{\dagger}_1\right)^2 = 0
,
\label{char1}
\ee
where $^\dagger$ stands for the operation of hermitian conjugation 
(as well as we employ the differential operators 
with real coefficients the hermitian
conjugation is equivalent to the operation of transposition, 
$Q^\dagger = Q^t$) .

The intertwining relations introduced in \gl{intertw} result in the 
Supersymmetry for a Super-Hamiltonian $H$,
\be
[H, Q_1] = [H, Q^\dagger_1] = 0 . \label{susytr}
\ee
The SUSY algebra is completed by the appropriate decomposition of
the Super-Hamiltonian,
\be
\fl H = \{Q_1 ,Q^{\dagger}_1\} \,\, 
\Longleftrightarrow\,\, h^+ =q^+_1 q^-_1 = - \partial^2 +\chi^2 -\chi' ;
\quad h^- =q^-_1 q^+_1  = - \partial^2 +\chi^2 +\chi', 
\label{susyfac}
\ee
which is in line with the Schr\"odinger 
one-step factorization \cite{schr,inf}. The notation $\chi' \equiv
d\chi/dx$ has been employed.

At this stage, the super-potential $\chi$ is generated by zero-energy 
solutions of the Schr\"odinger equations (equivalently, 
zero-modes of supercharges 
$Q_1, Q^{\dagger}_1$),
\be
h^\pm \phi^\mp_{(0)} = 0 = q^\mp_1  \phi^\mp_{(0)}; \quad  
\phi^-_{(0)} (x) = \left(\phi^+_{(0)} (x)\right)^{-1} = 
\exp\left(- \int^x dy \chi(y)\right) . \label{zero1}
\ee
If  $\chi(x)$ is a non-singular function then the zero-modes 
$\phi^\mp_{(0)}$ are
nodeless. This leads to a non-negative physical spectrum in
agreement
with the SUSY algebra \gl{susyfac},
\be
\langle\psi| H | \psi\rangle = \langle Q_1\psi| Q_1\psi\rangle +  
\langle Q^{\dagger}_1\psi| Q^{\dagger}_1\psi\rangle \geq 0 , \label{posit}
\ee
for any $L^2$-normalizable, smooth wave function $\psi (x)$.
  
Several options exist for the  choice  of zero-modes of supercharges
$Q_1, Q^{\dagger}_1$.
If one of the zero-modes  $\phi^\mp_{(0)}$ is normalizable then it becomes
a ground state wave function of the Super-Hamiltonian $H$ ({\it i.e.} 
of $h^+$ or $h^-$). But another one remains non-normalizable due to 
Eq.\gl{zero1}. Thus either $q^-$ or $q^+$ deletes the ground state level 
of $h^+$ or $h^-$. When keeping in mind the spectral design program one can 
also interpret it conversely: if $q^-$ deletes the lowest level
of  $h^+$ converting it into   $h^-$ then $q^+$ creates a new level for  $h^-$ 
transforming it into  $h^+$. 

Another option is realized by the non-normalizable nodeless 
functions  $\phi^\mp_{(0)}$ when none of them 
belongs to the physical spectrum of Hamiltonians  $h^\pm$. In this case 
the entire sets of physical eigenstates of the both Hamiltonians are 
put into the one-to-one correspondence by intertwining relations \gl{intertw},
\be
h^\pm \psi^\pm_E = E  \psi^\pm_E;\quad E > 0;\quad
 \psi^\mp_E =\frac{1}{\sqrt{E}} 
q^\mp_1 \psi^\pm_E,
\ee
and such Hamiltonians are strictly isospectral. In the SUSY vocabulary
it is the case of ``spontaneous'' SUSY breaking as the lowest ground state
of the Super-Hamiltonian $H$ is degenerate.

The previous analysis has been based not only on the intertwining relations
\gl{intertw} but also on the factorization \gl{susyfac}. However 
there is no such factorization for higher-order intertwining operators 
\gl{crum}.
What do we have for the latter ones instead?

\section{From the ladder of SUSY's via Parasupersymmetry 
toward Polynomial SUSY}

\hspace*{3ex} Let us proceed by recursion and 
discover different levels of isospectrality: 
from a simple Darboux transformation 
to a ladder or a dressing chain made of several simple Darboux steps.
Actually the whole variety of elementary building blocks 
for spectral design can
be well developed within the class of
 transformation operators  $q^{\pm}_2$ of second-order in derivatives.
One has to select the operators \gl{crum} 
with nonsingular coefficient functions
which produce a nonsingular potential $V_2$ after intertwining  \gl{intertw} 
with the smooth
initial potential $V_1$.

First of all,  to produce the required
transformation operators the two different linear SUSY systems
may be ``glued''. Indeed, consider two Super-Hamiltonians 
\mbox{$H_i, \ i = 1,2$}, Eq.~\gl{hamil}, respectively
two supercharges $Q_i$ with super-potentials $\chi_i$ and 
supercharge components \mbox{$r_i^\pm = \mp\partial + \chi_i $}. 
Let us identify two elements of Super-Hamiltonians,
 \be
h^-_1 = h^+_2 +\lambda;\quad \chi^2_1 +\chi'_1= \chi^2_2 -\chi'_2 +\lambda 
\label{glue1}
\ee
with $ E_1^{(0)}\geq \lambda \geq  - E_2^{(0)}$ where  $E_1^{(0)}$ and 
 $E_2^{(0)}$ are ground state energies for $h^-_1$ and  $h^+_2$ respectively.
Evidently the constant shift of the Super-Hamiltonian 
$H_2 \rightarrow H_2 +\lambda$ does not break or change the supersymmetry.

After such a gluing the chain of intertwining relations \gl{intertw} 
can be assembled into the supersymmetry transformation 
$[H_{ps}, Q_{ps}] = [H_{ps}, Q^{\dagger}_{ps}] = 0$ of the combined 
Super-Hamiltonian $H_{ps}$ and the joint supercharges $Q_{ps},Q^{\dagger}_{ps}$,
\be
\fl
H_{ps} = \left(\begin{array}{ccc}
h^+_1 -{\lambda\over 2}& 0& 0\\
0 & h_1^-  -{\lambda\over 2}= h^+_2 +{\lambda\over 2} &0\\
0&0&  h^-_2 +{\lambda\over 2}
\end{array}\right);\,\,\,\,  
Q_{ps}=\left(\begin{array}{ccc}
 0 &  r^+_1&0\\
0 & 0 &r^+_2\\
0&0&0
\end{array}\right), \label{para1}
\ee
where we have shifted both Super-Hamiltonians symmetrically to simplify the
further algebra.
However these supercharges   $Q_{ps},Q^{\dagger}_{ps}$ are not anymore nilpotent of
order two and therefore do not mimic the Pauli principle. Still they are
nilpotent of order three, $Q_{ps}^3 = \left(Q^{\dagger}_{ps}\right)^3 = 0$, 
and therefore 
the states carrying these charges obey the para-statistical principle.
Thus we deal now with the Parasupersymmetry \cite{ruba}--\cite{plyupara}.
Furthermore the closure of superalgebra is 
not anymore given by Eq.\gl{susyfac}.
The lowest-order relation between the Super-Hamiltonian and the para-supercharges
 is trilinear,
\be
Q^{\dagger}_{ps}Q^2_{ps} + Q^2_{ps}Q^{\dagger}_{ps} + 
Q_{ps}Q^{\dagger}_{ps}Q_{ps} =
2 H_{ps} Q_{ps} . \label{para2}
\ee 
This quantum system reveals triple degeneracy of levels with the 
possible exception for two lowest states. We draw also the reader's 
attention to a possibility to treat both  intermediate
Hamiltonians $h^-_1$ and $h^+_2 +\lambda$ separately, in spite of their 
identification, thus  doubling of the 
Hilbert spaces spanned by their eigenfunctions. 
It leads to a model of ``Weak Supersymmetry'' with quadruple level
degeneracy 
\cite{smilga} which may be susceptible to prolongation onto a weak-SUSY 
field theory. 

Going back to the spectral design purposes,
the para-supersymmetric dynamics contains redundant information, namely,
about the intermediate Hamiltonian  $h^-_1 = h^+_2 +\lambda$. One
is, in fact, interested in the final Hamiltonian $h^-_2$ only
as 
produced from the initial one, $h^+_1$ by means of   
a second-order Darboux transformation,
\be
\fl h^+_1 r^+_1 r^+_2 = r^+_1 h^-_1 r^+_2  =  r^+_1 (h^+_2 + \lambda) r^+_2 =
 r^+_1 r^+_2  (h^-_2 + \lambda) ; \quad
r^-_2 r^-_1 h^+_1  = (h^-_2 + \lambda) r^-_2 r^-_1 . \label{intertw2}
\ee
Let us make a shortcut and define 
the two isospectral components $h^\pm$,
\ba
&&h^+ \equiv h^+_1  + \lambda_1 =  r^+_1 r^-_1 + \lambda_1;\quad   
h^- \equiv h^-_2 + \lambda_2 =  r^-_2 r^+_2 + \lambda_2;\label{def2}\\
&&r^-_1 r^+_1 + \lambda_1 =  r^+_2 r^-_2 + \lambda_2; \label{glue2}
\ea 
for the generalized Super-Hamiltonian \gl{hamil} where we have employed a more
general energy reference (shift by arbitrary $\lambda_{1,2}$). 
Evidently, $\lambda =  \lambda_2 - \lambda_1$ .
Then the intertwining relations \gl{intertw} are identical to Eq.~\gl{intertw2}
with $q^+_2 =  r^+_1 r^+_2$ and the supersymmetry  $[H, Q_2] = [H, Q^{\dagger}_2] 
= 0$ is generated by the conserved supercharges,
\be Q_2=\left(\begin{array}{cc}
 0 &  q^+_2\\
0 & 0 
\end{array}\right),\quad  Q_2^2 = \left(Q^{\dagger}_2\right)^2 = 0 . 
\label{char2}
\ee
In place of Eq.\gl{susyfac}, because of \gl{def2} the 
algebraic closure is given by,
\be
 \{Q_2 ,Q^{\dagger}_2\} =   \left(\begin{array}{cc}
 r^+_1 r^+_2 r^-_2 r^-_1 & 0 \\
0 &   r^-_2 r^-_1 r^+_1 r^+_2
\end{array}\right) =   (H - \lambda_1) (H - \lambda_2). \label{polyn2}
\ee
Thus we have obtained the second-order Polynomial SUSY algebra
\cite{ais}-\cite{fern96} as a concise
form of isospectral deformation of a potential system
 accomplished by a ladder  
\cite{coop}--\cite{fern}, \cite{inf}--\cite{berez} or a dressing chain 
\cite{shab1}--\cite{adler} of a couple of one-step 
Darboux transformations
or, equivalently, 
by a second-order Crum-Darboux intertwining \cite{crum,ander,bagsam1} or
by a blocking of two linear SUSY with partial overlapping of 
Super-Hamiltonians \cite{ais} \mbox{(``weak SUSY'' \cite{smilga})}, 
or by a tower of para-SUSY transformations \cite{ruba}--\cite{aisv-ps}.

In the modern SUSY vocabulary there are several synonyms for the higher-order
SUSY algebra: originally it was named as a  
Polynomial (or Higher-derivative) one \cite{ais,acdi95}, 
recently the title of $N$-fold SUSY has been suggested
\cite{ast00} and, at last, 
a more general term of Nonlinear SUSY has
been used \cite{kliplyu00} with  a certain reference to  nonlinear SUSY
algebra arising in the conformal QM \cite{ivanov}. In what follows we will
combine the first name and the last one depending on the structure of a
\mbox{superalgebra \cite{ansok}.}

This Polynomial SUSY keeps track of essential
spectral characteristics of the second-order SUSY 
(Crum-Darboux transformations) . Indeed,
the zero-modes of intertwining operators $q^\pm_2$ or, equivalently, 
the zero-modes of the hermitian supercharges $Q^+_2 = Q_2 + Q^{\dagger}_2;\,  
Q^-_2 =i (Q^{\dagger}_2 - Q_2)$, form the basis of a 
finite-dimensional representation
of the Super-Hamiltonian,  
\be
q^\pm_2 \phi^\pm_i (x) = 0 = q^\pm_2 h^\mp  \phi^\pm_i (x);\quad i = 1,2;\quad
h^\mp  \phi^\pm_i (x) = \sum_{j=1}^2 S^\mp_{ij}   \phi^\pm_j (x), \label{zerom1}
\ee
due to intertwining relations \gl{intertw}, \gl{intertw2}.
In terms of these Hamiltonian projections -- constant matrices ${\bf S}^\mp$ ,
the SUSY algebra closure 
takes the polynomial form \cite{ansok} (see also \cite{ast00}),
\be
\left\{Q_2, Q^{\dagger}_2\right\} =
{\rm det} \left[E{\bf I}-{\bf S}^{+}\right]_{E = H}
={\rm det} \left[E{\bf I}-{\bf S}^{-}\right]_{E = H} \equiv
{\cal P}_2 (H) . \label{clos2}
\ee
Thus both matrices  $ {\bf S}^\mp$
 have the same set of eigenvalues which for the ladder
construction \gl{polyn2} consists of $\lambda_1, \lambda_2$ . As
the zero-mode set is not uniquely derived from \gl{zerom1} the matrices
$ {\bf S}^\mp$ are not necessarily diagonal. For instance, the equation
$r^+_1 r^+_2 \phi^+ (x) = 0$ has  one zero-mode $\phi^+_2$ obeying
$r^+_2 \phi^+_2 (x) = 0$ and another one obeying 
\mbox{$r^+_1 \tilde\phi^+_1 = 0;\, 
 \tilde\phi^+_1 = r^+_2 \phi^+_1 (x) \not= 0$.} 
Evidently the zero-mode solution
 $\phi^+_1 (x)$ is determined up to an arbitrary admixture of  $\phi^+_2$. When multiplying
these linear equations by $r^-_2$ one easily proves with the help of 
Eqs.~\gl{def2},\gl{glue2} that
\be
\fl (h^- - \lambda_2) \phi^+_2 (x) = 0;\quad (h^- - \lambda_1) \phi^+_1 (x)
= C  \phi^+_2 (x);\quad {\bf S}^- = \left(\begin{array}{cc}
 \lambda_1 &  C\\
0 & \lambda_2
\end{array}\right), \label{cellgen}
\ee
where $C$ is an arbitrary real constant. If $\lambda_1 \not= \lambda_2$ 
then by the redefinition \mbox{$(\lambda_1 -\lambda_2) \bar\phi^+_1 \equiv
(\lambda_1 -\lambda_2) \phi^+_1 +  C  \phi^+_2 (x)$} one arrives at 
the canonical diagonal form $\widetilde{\bf S}^-$. 
However in the {\sl confluent case},  
$\lambda_1 = \lambda_2 \equiv \lambda,\, C\not=0$ 
it is impossible to diagonalize 
and by  a proper normalization of the zero-mode  $\phi^+_1$
one gets another canonical form  $\widetilde{\bf S}^-$ of this matrix -- the 
elementary Jordan cell \cite{gant},
\be
\fl (h^- - \lambda) \phi^+_2 (x) = 0;\quad (h^- - \lambda) \phi^+_1 (x)
=   \phi^+_2 (x);\quad \widetilde{\bf S}^- = \left(\begin{array}{cc}
 \lambda &  1\\
0 & \lambda
\end{array}\right). \label{cell2}
\ee
We display it to emphasize that in the confluent case 
the zero-mode  $\phi^+_1$ is not anymore a solution of the Schr\"odinger 
equation but it is a so-called adjoint solution \cite{naim}  which can be obtained 
by differentiation,  $\phi^+_1
=d\phi^+_2/d\lambda$ . 
Yet the 
intermediate Hamiltonian $\tilde h =r^-_1 r^+_1 + \lambda 
= r^+_2 r^-_2 + \lambda $ is well defined and therefore 
the intermediate isospectral partner  $\tilde\phi^+_1 (x)$
of the zero-mode $\phi^+_1 (x)$ is a solution of Schr\"odinger equation 
with the above Hamiltonian.
The analysis of the matrix  ${\bf S}^+$ is similar.
Thus we have established that in general the Hamiltonian projection onto the
subspace of (hermitian) supercharge zero-modes is not diagonalizable but can
be always transformed into a canonical Jordan form. 

To accomplish the description of 
Polynomial SUSY algebras generated by a second-order
ladder one should take into consideration also the degenerate case
when \mbox{$\lambda_1 = \lambda_2 \equiv \lambda,\, C=0$}. For this choice
the matrix   ${\bf S}^-$ is automatically diagonal and both zero modes  
$\phi^+_{1,2} (x)$ are 
(independent) solutions of the   Schr\"odinger 
equation with  the  Hamiltonian  $h^-$. Then it can be proved \cite{ansok} that
the intertwining operator $q^+_2$ is just a linear 
function of this Hamiltonian,
$q^+_2 = \lambda - h^-$. Hence the intertwining is trivial $h^- = h^+$ and 
such supercharges must be eliminated. For higher-order SUSY the
removal of such blocks in supercharges 
 may lead to ladder irreducible SUSY algebras \mbox{(see Section 4).}

The very fact that the Hamiltonian is represented by finite matrices
 ${\bf S}^\pm$ is interpreted sometimes \cite{anst,sataka} 
as a phenomenon of quasi-exact 
solvability (QES) \cite{tur,shif}: this point needs a certain comment.
If one is seeking for some formal solutions of the Schr\"odinger equation,
not necessarily normalizable and regular then such a formal QES
can be accepted. But for the spectral design we impose physical boundary
conditions and requirements of normalizability which are essential to
define the energy levels properly. Then QES for 
physical wave functions is achieved only if one or both of eigenvalues
belong to the energy spectrum of the Super-Hamiltonian. Obviously it is an
exceptional situation which is not granted by the Polynomial SUSY itself.

Let us complete this section with the general description of the $N$-step 
ladder which entails the Polynomial Superalgebra of $N$th-order. We 
introduce a set of first-order differential
operators for intermediate intertwinings ,
\be
r^\pm_l = \mp \partial + \chi_l(x),
\quad l = 1,...,N, 
\ee
and the relevant number of 
intermediate super potentials $\chi_l(x)$. The set of the 
initial, $h^+ \equiv h_0$,
the final, $h^- \equiv h_N$ and intermediate Hamiltonians,
$ h_l = - \partial^2 + v_l (x)$ consists of Schr\"odinger operators, so far
nonsingular and real ones. They obey the
ladder relations (``gluing''),
\ba
 &&h_l \equiv r^-_l\cdot r^+_l  + \lambda_l =  r^+_{l+1}\cdot
r^-_{l+1} +
\lambda_{l+1} ,\quad l = 1,\ldots,N-1,\nonumber\\
&& h_N \equiv h^- = r^-_{N} \cdot r^+_{N} + \lambda_{N},\quad
 h_0 \equiv h^+ = r^+_1 \cdot r^-_1  + \lambda_1 . \label{ladderN}
\ea
These gluing relations are provided by the (dressing) chain equations on 
super-potentials,
\be  v_l (x) =
(\chi_l(x))^2 +
(\chi_l(x))' + \lambda_l=
(\chi_{l+1}(x))^2 -
(\chi_{l+1}(x))' + \lambda_{l+1}
\ee
The corresponding intertwining (Darboux) transformations hold for each
adjacent pair of Hamiltonians,
\be
h_{l-1} \cdot r^+_l =  r^+_l\cdot h_{l},\quad  r^-_l\cdot h_{l-1} =
h_{l} \cdot r^-_l,
\ee
and therefore the chain of $N$ overlapping SUSY systems is properly built,
\ba
&&H_l = \left(\begin{array}{cc}
h_{l-1}& 0\\
0 & h_l
\end{array}\right),\qquad  R_l=\left(\begin{array}{cc}
 0 &  r^+_l\\
0 & 0
\end{array}\right);\nonumber\\
&&[H_l, R_l] = [H_l, R_l^\dagger] = 0,\qquad
H_l - \lambda_l= \{R_l,R_l^\dagger\},
\ea
This Chain Supersymmetry can be equally converted into a $N$-order 
Parasupersymmetry similar to Eqs.~\gl{para1}, \gl{para2} which however we do
not need for the further construction.

Now let us disregard a chain of intermediate Hamiltonians between
  $h^+$ and $h^-$ and proceed to the Higher-derivative $\simeq$ 
Polynomial  
$\simeq$ Nonlinear SUSY  algebra for the Super-Hamiltonian $H$ given in 
Eq.\gl{hamil}. The intertwining between  $h^+$ and $h^-$ is realized by
the Crum-Darboux operators,
\be
q^+_N = r^+_1 \ldots r^+_N, \quad q^-_N = r^-_N \ldots r^-_1 . \label{inop}
\ee 
The SUSY symmetry
$[H,Q_N] = [H,Q^{\dagger}_N] =0, $ is still performed 
by the supercharges of the 
same matrix structure \gl{char2} and the Super-Hamiltonian is
represented by finite-dimensional matrices on the subspaces of
supercharge zero-modes,
\be
q^\pm_N \phi^\pm_i (x) = 0;\quad i = 1,2,\ldots,N;\quad
h^\mp  \phi^\pm_i (x) = \sum_{j=1}^N S^\mp_{ij}   \phi^\pm_j (x), 
\label{zeromN}
\ee
due to intertwining relations \gl{intertw}.
In terms of the constant matrices $ {\bf S}^\mp$ ,
the algebraic closure is just represented by  a non-linear SUSY
relation 
\cite{ansok, ast00},
\be
\fl \left\{Q_N, Q^{\dagger}_N\right\} =
{\rm det} \left[E{\bf I}-{\bf S}^{+}\right]_{E = H}
={\rm det} \left[E{\bf I}-{\bf S}^{-}\right]_{E = H} \equiv
{\cal P}_N (H) = \prod^N_{l=1} \left(H - \lambda_l\right) . \label{polialg}
\ee
The disposition of the real roots $\lambda_l$ against the energy levels of the
Super-Hamiltonian is assumed to provide the positivity of the superalgebra
\gl{polialg} (see next Section). Namely if $m$ lowest energy levels, 
$\{E_{j}\},\, j=0,\ldots,m-1$ are among the roots, $\{E_{j}\}\subset 
\{\lambda_l\}$ then all   
$\lambda_l < E_m$.
Again both matrices $ {\bf S}^\mp$ 
have the same set of eigenvalues which for the ladder
construction \gl{ladderN} consists of $\lambda_1,\ldots \lambda_N$ . 
If the degenerate roots appear then the canonical forms
$\widetilde {\bf S}^\mp$ of the latter matrices are not necessarily 
diagonal and may consist of Jordan cells.
If all intermediate $h_l$ 
are hermitian, nonsingular and superpotentials are taken real, 
then $\lambda_l$ are real and each ladder
step is well defined. What will happen if we extend the class of Polynomial
SUSY algebras admitting  complex $\lambda_l$  and singular $h_l$ ? 
\section{ Algebraic classification of Polynomial SUSY QM and its functional
realization: irreducibility of type I, II, and III}

\hspace*{3ex} Let us now examine which  circumstances may obstruct the SUSY ladder 
decomposition of a Polynomial SUSY algebra. In fact, all elementary "bricks"
irreducible to a chain of one-step Darboux intertwinings are well revealed
for the second-order SUSY algebra described in the previous section.

For a supercharge of second order in derivatives with real 
coefficient functions one can find real 
 zero-modes of the intertwining operators $q^\pm_2$ and, further on,
 the $2\times2$  matrix representation \gl{zerom1} for the 
Super-Hamiltonian components $h^\pm$ by matrices ${\bf S}^\pm$. 
The latter matrices
are real but, in general, not symmetric. Therefore the first obstruction 
for the ladder decomposition may arise because the reduction to a 
Jordan form has not given real eigenvalues. For instance,
if $h^+  \phi^-_i (x) = \bar\lambda \epsilon_{ik} \phi^-_i (x)$ then 
the eigenvalues of  ${\bf S}^+ =  \bar\lambda \hat\epsilon$ are imaginary, 
mutually conjugated $\pm i \bar\lambda$.  The possibility of complex pairs of 
mutually conjugated roots in a Polynomial SUSY algebra 
can be easily read off  from its closure \gl{clos2} 
as for supercharges with real coefficients
polynomials ${\cal P}_2 (H)$ possess real coefficients.
We call this kind of irreducibility to be of type I.
Its elementary block corresponds to the polynomial  
${\cal P}^{(I)}_2 (H) = (H + a)^2 + d,\quad d > 0$ and its analytical 
properties have been investigated in \cite{acdi95}, \cite{Fern:2002}.
Some examples of related isospectral potentials are described 
in \cite{deber2002}. 

Next, one has to ensure the positivity of the SUSY algebra relation \gl{clos2}
in a particular differential realization of a 
Super-Hamiltonian $H$  
with real non-singular potentials 
and the supercharges $Q_N, Q^{\dagger}_N$ (with $N=2$ in our case) 
made of differential operators with real
coefficients.
Let the energy spectrum $E_j; j=0,1,\ldots; E_j < E_{j +1}$ 
of $H$ be discrete, for simplicity. 
Then,
\be
 {\cal P}_N (E_j) = \langle Q_N\Psi_j | Q_N\Psi_j\rangle +
 \langle Q^{\dagger}_N\Psi_j | Q^{\dagger}_N\Psi_j\rangle \geq 0 , \label{ineq2} 
\ee
if the action of supercharges is well defined in the Hilbert space spanned
by eigenfunctions of a Super-Hamiltonian.
It can be easily extended on a continuum energy spectrum as well 
using wave packets.

Thus for regular potentials the allowed disposition 
of polynomial roots
  $\Longleftrightarrow$ zero-modes of a supercharge ---
provides non-negative values of  ${\cal P} (E)$ for {\sl each} energy
  level of a Hamiltonian. To be definite, one may have the following cases
for the allocation of polynomial real roots (for a pair of complex,
mutually conjugated roots the positivity is obvious) .
\smallskip

\noindent
{\it Case}\, A.\quad $\lambda_{1} \leq \lambda_2 \leq E_0$ or 
$\lambda_{1} = E_0;\, \lambda_2 = E_1 $.\\ The related SUSY algebra 
has a chain or ladder realization. In other words, it is
 reducible, in principle, because one gradually can 
add/remove $\lambda_1$ and then $\lambda_2$
without breaking the positivity of intermediate SUSY algebra.
The coincidence of roots and energies correspond in 
the spectral design to deleting/inserting energy levels. For instance,
if $\lambda_{1} = E_0;\, \lambda_2 = E_1 $ then two pairs of zero-modes 
of $q^\pm_2$ can be chosen as solutions of two Schr\"odinger equations
with Hamiltonians $h^\mp$. Repeating the arguments of Section 1 one can 
conclude \cite{suku1} that the energy levels $E_0, E_1$ may well appear in
any of the Hamiltonians $h^\pm$ but each level only once, either in $h^+$ or
in $h^-$. Thus the entire variety of spectral design tools happens to be 
at our disposal: namely, 
one may delete two lowest levels, replace the ground-state level
by a different one and add two more levels below the ground-state one.
\smallskip

\noindent
{\it Case}\, B.\quad  $E_0 < \lambda_{1} < \lambda_2 \leq E_1$  or  
$E_j \leq \lambda_{1} < \lambda_2 \leq E_{j+1},\, 1\leq j$.\\ A pair of real
roots is placed between adjacent energy levels. 
If one of the solutions with eigenvalues $\lambda_{1,2}$ (zero-modes of
the supercharges) is normalizable we perform the insertion/deletion of
an excited energy state. Thus with these means one can  delete two adjacent 
excited levels, shift the position of an excited level  and 
add two more excited  levels between two adjacent ones. 
Evidently, such an algebra
cannot be safely decomposed into a chain of two linear SUSY as the removal of
any of roots $\lambda_{1,2}$ immediately breaks the positivity in \gl{ineq2}.
Then the intermediate Hamiltonian  
acquires inevitably a real but singular potential leading 
to the loss of isospectrality.
The related Darboux transformations had been known in 50ties \cite{krein}.
We call this irreducibility to be of type II.
Examples and certain theorems are given in \cite{samsirr1},
\cite{deber2002}.
\smallskip

\noindent
{\it Case}\, C.\quad
$E_j < \lambda_{1} = \lambda_2 \leq E_{j+1},\, 0\leq j$.\\ This is a confluent
case which seems to arise as a limit of the previous one. However, let us 
remind that the one-dimensional QM does not allow degenerate levels. 
Besides, let's assume 
that the matrix representation for the corresponding Super-Hamiltonian
contains a non-trivial Jordan cell. Then the limit becomes quite delicate
as one of the zero-modes is not a solution of the Schr\"odinger 
equation but represents an adjoint function \cite{naim} 
(see Eq.~\gl{cell2} and 
discussion afterwards). This is why 
we specify this case as a separate one named to be of type III. 
With such an intertwining
operation one may insert/delete odd number of excited 
levels in an economical way. One may find
more information on the analytical properties of related potentials in 
\cite{Ferna:2003}.

One can  apply these second-order blocks and build an $N$th-order
Polynomial SUSY system. Their general form is again given by
Eq.~\gl{polialg} allowing the presence of complex conjugated roots
$\lambda_l$.
Let us rewrite it taking into account the possibility of complex and
degenerate roots,
\ba
&&\left\{Q_N,Q^{\dagger}_N\right\} = 
{\cal P}_N (H) = \prod^n_{l=1} \left(H - \lambda_l\right)^{\nu_l}
\prod^m_{j=1} [(H + a_j)^2 + d_j]^{\mu_j},\nonumber\\
&&N = \sum^n_{l=1} \nu_l + 2 \sum^m_{j=1} \mu_j,\quad d_j >0. \label{fullalg}
\ea
We stress that in the general case the Hamiltonian projections onto the
zero-mode spaces of intertwining operators $q^\pm_N$ are   finite
$N\times N$ matrices ${\bf S}^\mp$ and the Polynomial SUSY algebra can
be
represented by Eqs.~\gl{polialg}. 
Inequality \gl{ineq2} is certainly valid for
${\cal P}_N (H)$.

Irreducible elements of type II are not straightforwardly  
seen in the structure of the
Polynomial SUSY algebra and can be unraveled only after the
inspection of disposition of polynomial roots in respect to energy
levels. They fill the chain of intertwining 
operators being even order in derivatives and placing a
pair (or few pairs) of real roots  
$\lambda_{l} < 
\cdots < \lambda_{l+2k-1}$ (supercharge 
zero-modes) between two successive energy
levels $E_j < \lambda_{l} <\cdots< \lambda_{l+2k-1} < E_{j+r};\, r-1 \leq 2k$ if the 
intermediate levels $ \{E_{j+1},\ldots,E_{j+r-1}\}\subset \{\lambda_{l},
\ldots, \lambda_{l+2k-1}\};\,  r-1 \leq 2k$. The eigenvalues $E_j, E_{j+r}$
are assumed not to coincide with any of polynomial roots. Then the polynomial
\be
{\cal P}_{2k,j,r}^{({\rm II})} (H) = \prod^{2k-1}_{i=0}(H -  \lambda_{l+i});\quad 
{\cal P}_{2k,j,r}^{({\rm II})} (E_m) > 0 . \label{blockII}
\ee 
When the related zero-modes coincide with some eigenfunctions of
the Super-Hamiltonian the pertinent 
supercharges create or annihilate particular excited states in the components
$h^\pm$ of the Super-Hamiltonian.

Irreducible elements of type III fill the chain of Darboux transformations 
being represented by
even-order intertwining 
operators responsible for allocation of even number of real confluent
roots 
\mbox{ 
$\lambda_{l} = \lambda_{l+1}= \cdots = \lambda_{l+2\nu_l -1} $} (= supercharge 
zero-modes) between two adjacent energy
levels $E_j <\lambda_{l}   
\leq E_{j+1}$ for some $0\leq j$,
\be
{\cal P}_{2\nu_l,j}^{({\rm III})} (H) = (H -  \lambda_{l})^{2\nu_l} ;\quad 
{\cal P}_{2\nu_l,j}^{({\rm III})} (E_m) > 0 .\label{blockIII}
\ee  
No more than two zero-modes may be  solutions of  the Super-Schr\"odinger
equation, in particular, 
eigenfunctions
of the Super-Hamiltonian if  $E_j = \lambda_{l}$.  Other 
zero-modes are adjoint functions \cite{naim} to
a solution of the Schr\"odinger equation.

Finally, in general, the polynomial in Eq.~\gl{fullalg} can be  
factorized into the product of the elements \gl{blockII} and \gl{blockIII}
with roots located between two successive or adjacent levels. 
The related ladder of 
 Darboux transformations consists of reducible steps as well as of few 
irreducible elements of type I, II, III displayed in Eqs.~\gl{fullalg},
\gl{blockII}, \gl{blockIII}.

Yet the open question remains whether any irreducible element of
type II or III (\gl{blockII} or \gl{blockIII}) 
can decomposed into the ladder of
second-order
irreducible blocks with regular hermitian intermediate Hamiltonians
between them
in the ladder.
We are informed that essential progress in this direction has been
made by A.~V.~Sokolov and hope to see it published soon.

On the other hand, an experienced SUSY designer may be somewhat  
puzzled with the
very existence of irreducible super-transformations. Indeed
it is quite conceivable that a pair of supercharge zero-modes or even
a pair of new excited energy levels of the Super-Hamiltonian
can be inserted by successive application
of first-order intertwining (super) transformations between regular
Hamiltonians following the ladder algorithm described in the previous
Section. But the order of
the relevant ladder of first-order transformations and
respectively of the final Polynomial SUSY will be evidently higher
than two.
We come to the problem of possible relationship between first-order reducible
and irreducible SUSY algebras having the same Super-Hamiltonian.

The related important question concerns the degenerate roots. By general
arguments these roots  are distributed between different Jordan
cells in the canonical forms  
$\widetilde{\bf S}^\pm$ of the matrices ${\bf S}^\pm$. 
One can inquire on how many Jordan
cells
may coexist and if several cells appear then what is their role in the
supercharge structure. All these problems are clarified 
with the help of the Strip-off theorem \cite{ansok}.

\section{From reducible SUSY to irreducible one when 
equipped by the Strip-off theorem}

\hspace*{3ex} Let first elucidate the possible redundancy in supercharges which can
be eliminated without any changes in the Super-Hamiltonian ({\it i.e.} 
preserving the same potentials).
There exists a trivial possibility when the intertwining operators
$q^\pm_N$ and
$p^\pm_{N_1}$  for $N > N_1$ are related by a polynomial factor $F(x)$
depending on
the Hamiltonian,
\be
q^\pm_N =
F (h^\pm) p^\pm_{N_1}
=  p^\pm_{N_1}  F (h^\mp). \la{triv}
\ee
Obviously in
this case the appearance
of the second supercharge does not result in any new restrictions on potentials.

Thus the problem arises of how to separate the nontrivial part of a
supercharge and avoid numerous SUSY algebras generated by means of ``dressing''
\gl{triv}. It can be systematically realized with the help of the
following\\

\noindent
{\it ``Strip-off'' theorem}.
\smallskip

Let's admit the
construction given by Eqs.~\gl{zeromN} and \gl{polialg}. Then\\
a) the matrix ${\bf S}^-$ (or ${\bf S}^+$) generated by the
Hamiltonians
$h^-$ (or $h^+$) on the
subspace of zero-modes of the operator $q^+_N$ (or $q^-_N$), after 
reduction into the  Jordan form $\widetilde{\bf S}^-$ (or $\widetilde{\bf S}^+$),
may contain only one or two Jordan
cells with equal
eigenvalues  $\lambda_l$;\\
b) assume that there are $n$ pairs (and no more) of Jordan
cells with equal
eigenvalues  and with
 the sizes $\nu_l$ of a smallest cell in the $l$-th pair;\,\,
then this condition is necessary and sufficient to ensure for
the intertwining operator $q^+_N$ (or $q^-_N$) to be represented in the
factorized form:
\be
q^\pm_{N} = p^\pm_{N_1} \prod^n_{l=1} (\lambda_l - h^\mp)^{\nu_l}, \la{factor}
\ee
where $p^\pm_{N_1}$
are intertwining operators
of order $N_1 = N - 2 \sum^n_{l=1}\nu_l $
which cannot be  decomposed
further on in the product similar to \gl{triv} with $F (x) \not= const$.

\bigskip

\underline{Remark.} The matrices  $\widetilde{\bf S}^\pm$ cannot contain more
than two Jordan cells with the same eigenvalue $\lambda$ because otherwise the
operator $\lambda - h^\pm$ would have more than two linearly independent
zero-modes (not necessarily normalizable).\\
The full proof of this theorem  has been performed in \cite{ansok}.

Let us  illustrate the  Theorem in the \underline{Example:}\\
the matrix ${\bf S}^-$ for the intertwining operator $q^+_3$
with Jordan cells of different size having
the same eigenvalue. It is generated by the operators,
\be
p^\pm = \mp\partial + \chi,\quad h^\pm =p^\pm p^\mp + \lambda,\quad
q^+_3 = - p^+ p^- p^+ =p^+ (\lambda - h^-).
\ee
Respectively:
\be
\fl \begin{array}{cccc}
\phi^+_1:\,& p^- p^+\phi^+_1 = \phi^+_2 &\longrightarrow&
h^- \phi^+_1 = \lambda \phi^+_1 + \phi^+_2;\\
\phi^+_2:\,& p^+\phi^+_2 = 0 &\longrightarrow& h^- \phi^+_2 = \lambda \phi^+_2;\\
\phi^+_3:\,& p^+\phi^+_3 \not= 0,\, p^- p^+\phi^+_3 = 0&
\longrightarrow &h^- \phi^+_3 = \lambda \phi^+_3;
\end{array}
\quad
{\bf S}^- =\left(\begin{array}{ccc}
\lambda&1&0\\
0&\lambda&0\\
0&0&\lambda
\end{array}\right).
\ee

As a consequence of the ``Strip-off'' Theorem one finds that
the supercharge components cannot be factorized
in the form \gl{triv}
if the polynomial $\tilde{\cal P}_N (x)$ in the SUSY algebra closure 
\gl{polialg}
does not have the degenerate zeroes. Indeed the SUSY algebra closure
contains the square of polynomial $F(x)$, for instance,
\be
q^-_N q^+_N = F (h^-) p^-_{N_1} p^+_{N_1}  F(h^-)
=  F^2 (h^-) \tilde{\cal P}_{N_1} (h^-),
\la{dzero}
\ee
where $\tilde{\cal P}_{N_1} (x)$ is a polynomial of
lower order, $N_1 \leq N-2$.
Therefore each zero of the polynomial $F (x)$ will produce a double zero
in the SUSY algebra provided by \gl{dzero}.

Thus the absence of degenerate zeroes
is sufficient to have supercharges without redundancy in the sense
of Eq.~\gl{triv}. However it is not necessary because the degenerate zeroes
may well arise in the confluent ladder construction giving
new pairs of isospectral potentials \ci{acdi95}.

Now we proceed to uncover the origin of irreducible, type-II and -III 
transformations based on the strip-off factorization. For clarity
let us consider an example of
irreducible SUSY of type II with supercharges of second order
in derivatives (see previous Section).  Suppose that it realizes
insertion of two new energy levels between the ground
and first excited states. Then three lowest energy
levels $E_0 < E_1 < E_2$ are of importance to study the relevant SUSY
systems: the ground state level 
is degenerate between SUSY partners $h^+$ and
$h^-$, {\it i.e.} $E^+_0 = E^-_0$ whereas the two excited levels are
present only in the spectrum of $h^-$.

One can use the ladder construction \gl{ladderN}--\gl{inop} 
to prepare the same
spectral pattern. For this purpose, intertwining operators
\gl{inop} 
of, at least, fourth order in derivatives must be employed.
Indeed, one can prescribe 
the ladder steps for $q^\pm_4$ as follows: start from a
pair of isospectral Hamiltonians with ground state energies $E_3$; 
generate the level $E_0$ in the Hamiltonian $h^+$ using the
intertwining operators $r^\pm_1$, then sequentially
create in the spectrum of $h^-$ the state with energy 
$E_2 < E_3$ by means of $r^\pm_2$, next the energy level $E_1 < E_2$ using
$r^\pm_3$ and finally the ground state with energy $E_0 < E_1$
exploiting $r^\pm_4$.  These elementary
steps are reflected in zero-modes of the 
intertwining operators $q^+_4 = r^+_1 r^+_2 r^+_3 r^+_4$ and 
 $q^-_4 = r^-_4 r^-_3 r^-_2 r^-_1$. Namely the ground state of $h^+$ is
a zero-mode of $r^-_1$ ({\it i.e.} of $q^-_4$) whereas the eigenstates of
$h^-$ corresponding to
 $E_0, E_1, E_2$  are annihilated by the product $r^+_2 r^+_3 r^+_4$
({\it i.e.} by $q^+_4$) according to Eq.~\gl{zeromN}. In particular,
the ground state of  $h^-$ is
a zero-mode of $r^+_4$. As the ground
state energies coincide for $h^\pm$ the Hamiltonian
projections on the $q^\pm_4$ zero-mode space -- the matrices 
${\bf S}^\pm$ are, in general, not diagonalizable but have one rank-two
Jordan cell each. Thus, for instance,
\be
{\bf S}^- =\left(\begin{array}{cccc}
E_0&0&0&C\\
0&E_2&0&0\\
0&0&E_1&0\\
0&0&0&E_0
\end{array}\right)\quad \Longrightarrow \quad  
\tilde{\bf S}^- =\left(\begin{array}{cccc}
E_0&C&0&0\\
0&E_0&0&0\\
0&0&E_2&0\\
0&0&0&E_1
\end{array}\right) ,  \label{cell4}
\ee 
where a non-zero constant $C$ can be normalized to $C=1$. 
The canonical Jordan form $\tilde {\bf S}^-$ in \gl{cell4} is achieved by means of 
re-factorization
\mbox{$q^+_4 = r^+_1 r^+_2 r^+_3 r^+_4 = r^+_1 \tilde r^+_2 \tilde r^+_3
 \tilde r^+_4$} so that the annihilation of ground state for $h^-$ is
associated now with $\tilde r^+_2$.
Respectively, the Polynomial
SUSY algebra shows up one degenerate root, 
\be
{\cal P}_4 (H) = (H- E_0)^2(H- E_1)(H- E_2) . \label{alg4}
\ee
The Strip-off theorem tells us that this fourth-order algebra
cannot be optimized to a lower-order one because  there is no
replication of roots in different Jordan cells of $\tilde {\bf
  S}^\pm$ matrices.  However one may perform fine-tuning of Darboux
transformation parameters to provide the constant $C = 0$ in \gl{cell4}. This
peculiar choice moves the SUSY system into the environment of the
Strip-off theorem as now two rank-one cells in \gl{cell4} contain the same
eigenvalue $E_0$ . The SUSY algebra is still given by Eq.~\gl{alg4} but
the intertwining operators reveal a redundancy,
\be
q^+_4 =  (E_0 - h^-)\, q^+_2.
\ee
By construction the 
left-hand side of this relation is fully factorizable in elementary
binomials $r^+_j$ with hermitian nonsingular intermediate Hamiltonians. 
But in the right-hand side the operator  $q^+_2 = \tilde r^+_3
 \tilde r^+_4$ does not admit a further
factorization with a nonsingular intermediate Hamiltonian
because after removal of the redundant factor $(h^- - E_0)$ such a
factorization is forbidden by the positivity of the SUSY algebra, 
Eq.~\gl{ineq2}. 

One can easily extrapolate the previous argumentation to the case of
additional
degeneracy of excited levels $E_1 = E_2$ to analyze the irreducible
SUSY of
type III. Thus
we reach the important conclusions that:\\
a) the factorization \gl{inop} of intertwining operators $q^\pm_N$ is
not unique and there exist options to have more reducible ladders and less
reducible ones with a larger number of singular intermediate Hamiltonians;\\
b) (some of) irreducible algebras of type-II and -III can be
identified with special cases of fully reducible ladder-type algebras
when the Hamiltonian projections ${\bf S}^\pm$ have an appropriate
number of pairs of Jordan cells with coinciding eigenvalues;\\
c) there are many (almost) isospectral systems with different pattern of
excited states which cannot be interrelated with the help of
irreducible Darboux transformations of type-II or -III but can be built
with the help of higher-order reducible SUSY ladder.

Yet one may substantially  gain effectiveness 
when the spectral design program
allows to apply the irreducible transformations of type-II or -III in
order to embed a couple of energy levels between two excited
ones. Thus
a more rigorous investigation of the relationship
between the reducible and irreducible intertwinings is
welcome. Especially important is the proof that any type-II, -III
irreducible SUSY can be embedded (for the same Super-Hamiltonian) into
a reducible ladder SUSY.

\section{More supercharges $\Longleftrightarrow$ Extended SUSY
$\Longleftrightarrow$ Hidden symmetry}
\hspace*{3ex}
 The possibility of two supercharges for a quantum SUSY system
was  mentioned in
\cite{acin2000} (see the preprint version). Namely
the conserved
supercharges $Q, Q^{\dagger}$ with complex coefficient functions in intertwining
components $q^\pm_N$ accounted for
two SUSY algebras for a hermitian Super-Hamiltonian $H$:
 one for their ``real'' parts $K,  K^\dagger $  and another one
for their ``imaginary'' parts $P, P^\dagger$
where 
real and imaginary parts are referred to
coefficients
in the intertwining operators
$q^\pm_N = k^\pm_{N} + i p^\pm_{N_1}$ .

Let us examine the general possibility to have
several supercharges for the same Super-Hamiltonian. First we remind
that a number of supercharges can be produced with the help of
multiplication
by a polynomial of the Hamiltonian (see Section 4). Certainly such
supersymmetries are absolutely equivalent for the purposes of spectral
design and one must get rid of them. As shown in \cite{ansok}
one can always optimize the infinite set of possible supercharges so
that no more than two nontrivial supercharges remain which are used to
generate all other ones by ``dressing'' with polynomials of the
Hamiltonians. 
Thus in one-dimensional QM one may have the ${\cal N} = 1,2$ SUSY only.

Correspondingly we consider now a general case when 
the Super-Hamiltonian $H$ admits two
supersymmetries with supercharges $K$ and $P$ generated by differential
intertwining operators of order $N$ and $N_1$ respectively,
\be
[H, K] = [H, P] =
 [H,  K^\dagger] = [H, P^\dagger]
= 0. \la{extsusy}
\ee
The second supercharge $P$
is assumed to be
a differential operator of lower order $N_1 < N$.

To close the algebra one has to include
all anti-commutators between supercharges, {\it i.e.}
the full algebra based on two supercharges $K$ and $P$
with real intertwining operators. Two supercharges generate
two Polynomial SUSY,
\be
\left\{K, K^\dagger \right\} = \tilde{\cal P}_N (H),\quad
\left\{P, P^\dagger \right\} = \tilde{\cal P}_{N_1} (H). \la{2alg}
\ee

The closure of the extended, ${\cal N} =2$ SUSY algebra is given by
\ba
\left\{P, K^\dagger \right\} & \equiv & {\cal R}
= \left(\begin{array}{cc}
 p^+_{N_1} k^-_{N} &  0\\
0 & k^-_{N} p^+_{N_1}
\end{array}\right),\no
\left\{K, P^\dagger\right\} &\equiv& {\cal R}^\dagger
= \left(\begin{array}{cc}
 k^+_{N} p^-_{N_1} &  0\\
0 & p^-_{N_1} k^+_{N}
\end{array}\right). \la{roper}
\ea
Apparently the
components of operators ${\cal R},\, {\cal R}^\dagger$
are differential
operators of $N + N_1$ order commuting with the Hamiltonians $h^\pm$, 
hence ${\cal R},\, {\cal R}^\dagger$ are symmetry operators  
for the Super-Hamiltonian.
 However, in general, they are not
polynomials of the Hamiltonians $h^\pm$ and these symmetries impose certain
constraints on potentials. 

All four operators
$\tilde{\cal P}_N (H),\,  \tilde{\cal P}_{N_1} (H),\, {\cal R},\, 
{\cal R}^\dagger$
mutually commute. Moreover the hermitian matrix describing this
 ${\cal N}=2$ SUSY,
\ba
{\cal Z} (H) = \left(\begin{array}{cc}
 \tilde{\cal P}_N (H) & {\cal R}  \\
{\cal R}^\dagger & \tilde{\cal P}_{N_1} (H)
\end{array}\right), \quad \mbox{\rm det} \left[{\cal Z} (H)\right] =
 \tilde{\cal P}_N \tilde{\cal P}_{N_1} -  {\cal R} {\cal R}^\dagger = 0, 
\la{centr}
\ea
is degenerate. Therefore it seems that the two supercharges are not
independent and by their redefinition
(unitary rotation) one might reduce the extended SUSY to an ordinary
${\cal N}=1$ one. However such rotations cannot be global and must
use non-polynomial, pseudo-differential operators as ``parameters''.
Indeed, the operator components of the ``central charge'' matrix ${\cal Z} (H)$
have different order in derivatives. Thus, globally the extended nonlinear
SUSY
deals with two sets of supercharges but when they act on
a given eigenfunction of the
Super-Hamiltonian $H$ one could, in principle, perform the energy-dependent
rotation and
eliminate a pair of supercharges. Nevertheless this reduction can be
possible only after the constraints on potentials have been resolved.

Let us find the formal relation between
the symmetry operators ${\cal R}, \,{\cal R}^\dagger$ 
and the Super-Hamiltonian.
These operators can be decomposed into a hermitian and an anti-hermitian
parts,
\be
{\cal B}\equiv \frac12({\cal R} + {\cal R})^\dagger \equiv \left(\begin{array}{cc}
b^+ & 0\\
0 & b^-
\end{array} \right),\quad
i {\cal E} \equiv \frac12 ({\cal R} - {\cal R}^\dagger)\equiv 
i \left(\begin{array}{cc}
e^+ & 0\\
0 & e^-
\end{array} \right). \la{herm}
\ee
The
operator ${\cal B}$ is a differential operator of even order and therefore
it may be a polynomial of the Super-Hamiltonian $H$. But if the
operator ${\cal E}$ does not vanish identically
it is a differential operator of {\it odd} order and
 cannot be realized by a polynomial of $H$.

The first operator
plays essential role in the one-parameter
non-uniqueness of the SUSY algebra.
Indeed, one can always redefine  the higher-order supercharge as
follows,
\be
 K^{(\zeta)} =  K + \zeta P,\quad
 \left\{ K^{(\zeta)}, K^{(\zeta)\dagger}\right\} =
\tilde{\cal P}^{(\zeta)}_N (H), \la{redef}
\ee
keeping the same order $N$ of Polynomial SUSY for arbitrary real
parameter $\zeta$. From \gl{redef} one gets,
\be
2 \zeta  {\cal B} (H) =
\tilde{\cal P}^{(\zeta)}_N (H) -\tilde{\cal P}_N (H) - \zeta^2
 \tilde{\cal P}_{N_1} (H),
\ee
thereby the hermitian operator ${\cal B}$ is a
polynomial of the Super-Hamiltonian of the order $N_b \leq N -1$.
Let's use it to unravel the Super-Hamiltonian content of the operator ${\cal E}$,
\be
{\cal E}^2 (H) = \tilde{\cal P}_N (H)
\tilde{\cal P}_{N_1} (H) - {\cal B}^2 (H), \la{secsym}
\ee
which follows directly from \gl{centr} and \gl{herm}. As
the (nontrivial) operator ${\cal E} (H)$
is a differential operator of odd order $N_e$ it may have only a
realization non-polynomial in $H$ being a square root of \gl{secsym} in
an operator sense. This operator is certainly non-trivial if the sum of orders
$N + N_1$ of the operators $k^\pm_N$ and $p^\pm_{N_1}$ is odd and therefore
$N_e = N + N_1$.  The opposite statement was also shown in
\cite{ansok}, 
namely if the symmetry
operator is non-zero then
for any choice of the operators $k^\pm_N$ and $p^\pm_{N_1}$ an optimal set of
independent supercharges (possibly of lower orders)
can be obtained which is
characterized by an odd sum of their orders.

The existence of a
nontrivial symmetry operator ${\cal E}$ commuting with the
Super-Hamiltonian results in common eigenstates which however are not necessarily
physical wave functions. In general they may be combinations of two solutions
of the Shr\"odinger equation with a given energy, the physical and
unphysical ones. But if  the
symmetry operator ${\cal E}$  is
hermitian in respect to the scalar product of the
Hilbert space spanned by the eigenfunctions of the Super-Hamiltonian $H$ then
both operators have a common set of physical wave functions. This fact imposes quite rigid
conditions on  potentials.

In particular,
for
intertwining operators with sufficiently smooth coefficient functions having
constant asymptotics for $x \longrightarrow \pm\infty$ the
symmetry operator ${\cal E}$  has the similar properties and is evidently
hermitian. In this case one has  non-singular
potentials with constant asymptotics  and therefore
a continuum energy spectrum  of $H$
with wave functions satisfying the scattering conditions.
Thus the incoming and outgoing states, $\psi_{in}(x)$ and $\psi_{out}(x)$,
at large $x$ are conventionally represented by  combinations of plane waves
which are solutions of the Schr\"odinger equation for a free particle,
\ba
&&\psi (x)|_{x \rightarrow -\infty} \longrightarrow
\exp(ik_{in}x) + R(k_{in})\exp(-ik_{in}x),\no
&&\psi(x)|_{x \rightarrow +\infty} \longrightarrow
\left(1 + T(k_{out})\right)\exp(ik_{out}x),
\ea
where the reflection, $R(k_{in})$, and transmission, $T(k_{out})$, coefficients
are introduced. Since the symmetry  is described by a differential operator
of odd order  which tends to an antisymmetric
operator with constant
coefficients the eigenstates of this operator 
approach asymptotically  individual plane waves
$\sim \exp(\pm ikx)$ with opposite eigenvalues $\sim \pm k f(k^2)$
and cannot be superimposed. Hence the eigenstate of the
Super-Hamiltonian with a given value of the operator ${\cal E}$ may characterize
only the transmission and cannot have any reflection, $R(k_{in}) = 0$.
We conclude that the corresponding partner potentials $V_{1,2}$
 inevitably belong to the
class of transparent or reflectionless ones \ci{refl}.
Such a
  symmetry may have relation to the Lax method in the soliton theory
\cite{matv}.

As the
symmetry operator ${\cal E}$  is
hermitian its eigenvalues are real but, by construction,
its coefficients are purely imaginary. Since the
wave functions of bound
states of the system $H$ can be always chosen real we conclude that they
must be zero-modes of the operator ${\cal E} (H)$,
\be
{\cal E} (H) \psi_i = {\cal E} (E_i) \psi_i = 0,\quad
  \tilde{\cal P}_N (E_i)\tilde{\cal P}_{N_1} (E_i) - {\cal B}^2 (E_i) = 0,
\la{zeroeq}
\ee
which represents the algebraic equation on bound state
energies of a system possessing two supersymmetries. Among solutions
of \gl{zeroeq} one reveals
also a zero-energy state at the bottom of
continuum spectrum. On the other hand
one could find also the solutions which are not
associated to any bound state. The very
appearance of such unphysical solutions is
accounted for by the trivial possibility to replicate supercharges by their
multiplication by the polynomials of the Super-Hamiltonian as discussed
in Sec.~4.
%%%%%%%%%%%%%%%%%%%
\section{A simple but useful example of Extended SUSY} 
Let us examine the algebraic structure of the simplest non-linear SUSY
with two supercharges,
\ba
k^\pm &\equiv & \partial^2 \mp 2f(x)\partial + \tilde b(x) \mp f'(x) ; \no
p^\pm &\equiv & \mp\partial + \chi(x), \la{exgen}
\ea
induced by the complex supercharge
of second order in derivatives \cite{ansok,acin2000}.
The supersymmetries \gl{extsusy}
generated by $K,\,  K^\dagger$ and $P, \, P^\dagger$ prescribe that
\ba
V_{1,2} &=& \chi^2 \mp \chi' =
\mp 2f' + f^2 + \frac{f''}{2f} - \left(\frac{f'}{2f}\right)^2 -
\frac{d}{4f^2} -a,\no
\tilde b &=& f^2 - \frac{f''}{2f} +\left(\frac{f'}{2f}\right)^2 +
\frac{d}{4f^2},
\ea
where $\chi, f$ are real functions and $a, d$ are real constants.
The related superalgebra closure for $K,\,  K^\dagger$ and $P, \, P^\dagger$
takes the form,
\be
\{K, \, K^\dagger\} = (H + a)^2 + d,\quad  \{P, \, P^\dagger\} = H.
\la{secor}
\ee
The compatibility of two supersymmetries is achieved by
 the following constraint $\chi = 2 f$ and by the
 nonlinear second-order differential equation
\be
f^2 + \frac{f''}{2f} -
\left(\frac{f'}{2f}\right)^2 - \frac{d}{4f^2} -a = \chi^2 =
 4 f^2. \la{2equ}
\ee
with solutions parameterized by two integration constants. Therefore
the existence of two SUSY reduces substantially
the class of potentials for which they may appear.
Evidently Eq.~\gl{2equ} can be integrated to,
\be
(f')^2 = 4 f^4 + 4 a f^2 + 4 G_0 f - d \equiv \Phi_4(f),
\la{firstor}
\ee
where $G_0$ is a real constant.

The solutions of this equation are elliptic functions which
 can be easily found in the implicit form,
\be
\int^{f(x)}_{f_0} \frac{df}{\sqrt{\Phi_4(f)}} = \pm (x - x_0),
\ee
where $f_0$ and $x_0$ are real constants.

It can be shown that they are nonsingular if:\\
a)  $\Phi_4(f)$ has three different real roots and the double root
$\beta/2$
is either the maximal one or a minimal one,
\be
 \Phi_4(f) = 4 (f - \frac{\beta}{2})^2
\left((f + \frac{\beta}{2})^2 - (\beta^2 -
\epsilon)\right),\quad
 0 < \epsilon < \beta^2. \la{polyn}
\ee
Then there exits a relation
between constants $a, d, G_0$ in terms of
coefficients $\beta,\epsilon$,
\be
a = \epsilon - {3 \beta^2\over 2} < 0,\quad G_0 = \beta
(\beta^2 - \epsilon),
\quad d = \beta^2  \left({3 \beta^2\over 4} - \epsilon \right) .
\la{defa}
\ee
The constant $f_0$ is
taken between the double root and a nearest simple root.\\
b)  $\Phi_4(f)$ has two different real double roots which corresponds
in \gl{polyn}, \gl{defa} to
$G_0 = 0,\quad \beta^2 = \epsilon > 0, \quad a = -\epsilon/2,\quad
d = -\epsilon^2/4$. The constant $f_0$ ranges between the roots.

The corresponding potentials $V_{1,2}$ are well known \ci{refl} to be
reflectionless, with one bound state at the energy $ (\beta^2 -
\epsilon)$ and
with the continuum spectrum starting from $ \beta^2$.
Respectively the scattering wave function is proportional to $\exp(ikx)$
with $k = \sqrt{E - \beta^2 }$.

In the case a) the potentials coincide in their form
and differ only  by shift in the coordinate (``Darboux Displacement''
\cite{Fernandez:2002wh}),
\be
V_{1,2} =  \beta^2 -
\frac{2\epsilon }{\mbox{\rm ch}^2 \left(\sqrt{\epsilon}(x -
x^{(1,2)}_0)\right)},\quad x^{(1,2)} =x_0 \pm \frac{1}{4\sqrt{\epsilon}}
\ln\frac{\beta - \sqrt{\epsilon}}{\beta + \sqrt{\epsilon}},
\la{caseb}
\ee
and  in the case b) one of the potentials can be chosen constant
(being a limit of infinite displacement),
\be
V_1 = \beta^2,\quad V_2 =  \beta^2 \left(1 -
\frac{2}{\mbox{\rm ch}^2 \left(\beta (x - x_0)\right)}\right),
\la{casec}
\ee
For these potentials one can elaborate extended SUSY algebra.

The initial algebra is given by the
relations \gl{secor}. It must be completed by the mixed anti-commutators
\be
\{ K,\,  P^\dagger\} = \{K^\dagger,\, P\}^\dagger =
{\cal B} (H) - i {\cal E} (H), \la{mixed}
\ee
where the first term is (see the next section) a polynomial of the
Super-Hamiltonian and 
the second one is in general not. In our case the first, polynomial
symmetry operator turns out to be constant, $ {\cal B} (H) = G_0$
when taking into account \gl{exgen} and \gl{firstor}.
Meanwhile the second
operator reads,
\be
{\cal E} (H) = i\left[{\bf I}\,\partial^3 - \Bigl(a {\bf I} + 
\frac32 {\bf V}(x)\Bigr) \partial 
- \frac34 {\bf V}' (x)\right],
\la{oddrep}
\ee
in the notations $
H \equiv -\partial^2 {\bf I} + {\bf V}(x)$. 
By construction the operator ${\cal E} (H)$ realizes a new symmetry
for the Super-Hamiltonian.
Directly from Eq.~\gl{oddrep}
one derives  that,
\be
{\cal E}^2 (H) = H \left[ (H + a)^2 + d\right]
 -  G_0^2  =  (H - E_b)^2 (H -  \beta^2),\la{quadr}
\ee
where $E_b = \beta^2 - \epsilon $ is the energy of a bound state.
Thus (some of) the zero modes of ${\cal E} (H)$ characterize either bound states or
zero-energy states in the continuum. However there exist also the 
non-normalizable, unphysical zero-modes corresponding to $E = E_b, \beta^2$. 
We remark that in the case b)
only the Hamiltonian $h^-$ has a bound state. Hence the physical 
zero modes of
 ${\cal E} (H)$ may be either degenerate [case a), broken SUSY] or 
 non-degenerate [case b), unbroken SUSY].

The square root of \gl{quadr} can be established unambiguously from the
analysis of transmission coefficients,
\be
{\cal E} (H) =  (H - E_b) \sqrt{H -  \beta^2}.\la{root}
\ee
We notice that 
the symmetry operator \gl{oddrep}, \gl{root} is irreducible, {\it i.e.}
 the binomial $ (H - E_b)$ cannot be ``stripped off''. Indeed the
elimination of this binomial would convert the third-order
differential
operator \gl{oddrep} into an essentially nonlocal
operator. The sign of square root in \gl{root}
is fixed from the conventional asymptotics
of scattering wave functions
$\sim \exp(ikx)$ and the asymptotics $V_{1,2} \longrightarrow \beta^2$
by comparison of this relation with Eq.~\gl{oddrep}.

When taking Eq.~\gl{root} into account one finds the mixed anti-commutators
of the extended  SUSY algebra \gl{mixed} in a non-polynomial form,
\be
\{ K, \, P^\dagger\} = \{K^\dagger,\, P\}^\dagger =
G_0 - i (H - E_b) \sqrt{H -  \beta^2}. \la{nonpol}
\ee
Thus the ``central charge'' of this extended SUSY is built of the elements
\gl{secor} and \gl{nonpol} and evidently cannot be diagonalized by a unitary
rotation with elements polynomial in $H$. Therefore the algebra must be
considered to be extended in the class of
differential operators of finite order.

Let us now clarify the non-uniqueness of the
higher-order supercharge and its role in the classification of the
Polynomial SUSY. For arbitrary $\zeta$ in \gl{redef} one obtains
\ba
&&\{K^{(\zeta)}, \, \left(K^{(\zeta)}\right)^\dagger\} 
= H^2 + (2a + \zeta^2) H + a^2 + d
+ 2 \zeta G_0 = (H + a_\zeta)^2  + d_\zeta,\no
&& a_\zeta = a + \frac12 \zeta^2,\quad
d_\zeta = d + 2 \zeta G_0 - a \zeta^2 - \frac14 \zeta^4
\equiv - \Phi_4 (- \frac{\zeta}{2}),
\ea
where $ \Phi_4 (f)$ is defined in Eq.~\gl{firstor}.

For the extended SUSY one can  discover that the previous
classification (Section 3) of
irreducible ladders (Darboux transformations) may fail. Indeed, 
the sign of $d_\zeta$, in general, depends on the choice of $\zeta$.
For instance, let us consider 
the case a) when
\be
d_\zeta  = - \frac14 \left(\zeta + \beta\right)^2
\left[ \left(\zeta - \beta \right)^2 - 4 (\beta^2 -\epsilon)\right].
\la{dlam}
\ee
Evidently if $\zeta$ lies between the real
roots of the last factor in \gl{dlam}
then $d_\zeta$ is positive and otherwise it is negative. But two real roots
always exist because $\beta^2 >\epsilon$.
Thereby the sign of $d_\zeta$
can be made negative as well as positive without
any change in the Super-Hamiltonian. Hence  in the case when
the Polynomial SUSY is an extended one,
with two sets of supercharges, the irreducibility of type I
of a Polynomial SUSY algebra does not signify any
invariant characteristic of potentials.

\section{Non-stationary Schr\"odinger equations: intertwining  and 
hidden symmetry}
\hspace*{3ex} Discussions of symmetry properties of the time dependent Schr\"odinger equation
have a long history, see for instance \cite{boyer} and references
quoted therein as well as \cite{okubo}. In these discussions the
potentials concerned are mainly time independent, see e.g. \cite{nikitin},
\cite{sergheyev}. Here our aim is to elucidate that many of the nonlinear SUSY
constructions illustrated before can be implemented also in the Schr\"odinger
time dependent framework \cite{bagsam1}, \cite{bagsa96}, \cite{Finkel98}.

Certainly, general first- and higher-order 
intertwining relations between non-stationary
one-dimensional Schr\"odinger operators can be easily  introduced. 
But already in the first-order
case the intertwining relations imply some hidden
symmetry which in turn leads to a specific quantum dynamics when the
evolution is described by quantum orbits and results 
in the $R$-separation of variables\cite{tdi}. 
Second-order intertwining operators \cite{tdi},\cite{Beckers:1998np}
and the corresponding non-linear SUSY  give rise to the quantum motion
governed by the spectrum generating algebras.

 Let us start with intertwining relations of the {\em
  non-stationary Schr\"odinger operator}
\begin{equation}
 {\cal S}[V]=i\partial_t+\partial^2_x - V(x,t)\;.
  \label{S}
\end{equation}
Here $\partial_t=\partial/\partial t$ and 
$\partial_x=\partial/\partial x$ denote the partial derivatives with respect
to time and position : we will denote these derivatives,
if applied to some function $f$, by a dot and prime, respectively. Hence, we
use the notation $\dot{f}(x,t)=(\partial_t f)(x,t)$ and $f'(x,t)=(\partial_x
f)(x,t)$.

The most general intertwining operator of first order \cite{tdi} is given by 
\begin{equation}
  \label{q+}
  q_t^+ =\xi_0(x,t)\partial_t + \xi_1(x,t)\partial_x + \xi_2(x,t)
\end{equation}
with, in general, complex-valued functions $\xi_0,\xi_1$ and $\xi_2$.The
possibility of a
complexification of the intertwining (Darboux) (first and also higher order)
operator
was emphasized by \cite{Beckers:1998np}. Note also
that in contrast to \cite{bagsa96} the formalism 
of \cite{tdi} allow a priori for a 
first-order operator in $\partial_t$.

For the above defined Schr\"odinger operator (\ref{S}) 
the intertwining relation reads
\begin{equation}
  \label{inS}
  {\cal S}[V_1]q_t^+ = q_t^+ {\cal S}[V_2]\;,
\end{equation}
where the functions $\xi_i$ ($i=0,1,2$) and  $V_{1,2}$
are  not independent. It can be also represented in the
SUSY form Eq.~\gl{susytr} when the stationary Hamiltonians $h^\pm$
are extended to the Schr\"odinger operators ${\cal S}[V_{1,2}]$, then
\be
[\hat{\cal S}_t, Q_t ] = 0, \quad  \hat{\cal S}_t = \left(\begin{array}{cc}
{\cal S}[V_{1}]& 0\\
0 & {\cal S}[V_{2}]
\end{array}\right),\quad  Q_t = \left(\begin{array}{cc}
 0& q_t^+\\
0 & 0
\end{array}\right). \label{susytime}
\ee
Inserting the explicit forms of the Schr\"odinger operators (\ref{S}) and the
intertwining operator (\ref{q+}) into relation (\ref{inS}) it was
found \cite{tdi} that 
$\xi_0$ and $\xi_1$  may depend only on time, i.e.\ $\xi_0'=0=\xi_1'$. The 
assumption
that $\xi_0$ does not vanish identically then leads to the consequence
that also the
potential difference $V_1-V_2$ does depend only on time. This is a rather
uninteresting case
and, therefore, we set $\xi_0\equiv 0$ without a loss of
generality. 
Making now the following choice of appropriate variables 
$\xi_1(t)=e^{i\beta(t)}\rho(t)$ and
$\xi_2(x,t)=e^{i\beta(t)}\rho(t)\omega'(x,t)$ with real $\beta$,
positive $\rho$ and  complex $\omega$ functions one finds
\begin{equation}
  \label{V1andV2:1}
  \begin{array}{l}
   V_1(x,t)=\omega'\,^2(x,t)+\omega''(x,t) - i \dot{\omega}(x,t) +
   \alpha(t)-\dot{\beta}(t)+i\dot{\rho}(t)/\rho(t)\;,\\[2mm]
   V_2(x,t)=\omega'\,^2(x,t)-\omega''(x,t) - i \dot{\omega}(x,t) + \alpha(t)\;,
  \end{array}
\end{equation}
where $\alpha$ is some time-dependent complex-valued integration constant. 
Again one may set \cite{tdi}
$\beta\equiv 0$ without loss of generality. Furthermore, one may also take
$\alpha\equiv 0$ because it can always be absorbed in $\omega$ by the
shift 
$\omega\to \omega -i\int dt\,\alpha$. Hence, we are left with 
\begin{equation}
  \label{V1andV2:2}
  \begin{array}{l}
   V_1(x,t)=\omega'\,^2(x,t)+\omega''(x,t) - i \dot{\omega}(x,t) +
   i\dot{\rho}(t)/\rho(t)\;,\\[2mm]
   V_2(x,t)=\omega'\,^2(x,t)-\omega''(x,t) - i \dot{\omega}(x,t)\;.
  \end{array}
\end{equation}
Here the so-called super-potential $\omega$ is still not arbitrary as the
potentials are to be real. This can, for example, be achieved
by assuming a stationary real super-potential. However it leads to
the standard stationary SUSY QM discussed previously. Alternatively, we will
consider a complex super-potential 
\begin{equation}
  \omega(x,t)= \omega_R(x,t) + i \omega_I(x,t)
\label{chi}
\end{equation}
with real functions $\omega_R$ and $\omega_I$. The reality condition
  $\mbox{Im }V_1=\mbox{Im }V_2=0$ implies
\begin{equation}
  \label{h_and_g}
  2(\omega_I)''+\dot{\rho}/\rho=0\;,\qquad 2(\omega_R)'(\omega_I)'-
(\omega_I)''-\dot{\omega_R}=0\;,
\end{equation}
which can easily be integrated to
\begin{equation}
  \begin{array}{rcl}
\omega_I(x,t)&=&\displaystyle
      -\frac{1}{4}\,\frac{\dot{\rho}(t)}{\rho(t)}\,x^2+
            \frac{1}{2}\,\rho(t)\dot{\mu}(t)x +\gamma(t)\;, \\
  \omega_R(x,t)&=&\displaystyle
      \frac{1}{2}\,\ln\rho(t)+K\Bigl(x/\rho(t)+\mu(t)\Bigr)\;,
  \end{array}\label{g_and_h}
\end{equation}
where $\mu$ and $\gamma$ are arbitrary real functions of time and $K$ is
an arbitrary real function of the variable $y=x/\rho +\mu$. In terms of
these functions the final form of the two partner potentials is
\begin{eqnarray}
\fl  V_{1,2}(x,t)=\frac{1}{\rho^{2}(t)}\left[K'^2(y)\pm K''(y)\right]
\nonumber\\-
  \frac{\ddot{\rho}(t)}{4\rho (t)}\,x^{2}+\left(\dot{\rho}(t)\dot{\mu}(t)+
  \frac{\rho(t)\ddot{\mu}(t)}{2}\right)x-\frac{\rho^2(t)\dot{\mu}^2(t)}{4}
  +\dot{\gamma}(t) 
\label{V_1/2}
\end{eqnarray}
and the intertwining operator reads
\begin{equation}
  q_t^ +(x,t)=\rho(t)\partial_x+K'\Bigl(x/\rho(t)+\mu(t)\Bigr)-
  \frac{i}{2}\left(\dot{\rho}(t)x-\rho^2(t)\dot{\mu}(t)\right).
\label{q+_2}
\end{equation}
Let us demonstrate \cite{tdi} that the non-stationary Schr\"odinger 
equation ${\cal S}[V_{1,2}]\psi_{1,2} =0$ 
with potentials given in Eq.~(\ref{V_1/2}) (which is equivalent to the
intertwining (\ref{inS})) admits a separation of variables. In
fact, after the transformation 
\begin{equation}
\label{trafo}
\fl  y=x/\rho(t) + \mu(t)\;,\qquad 
  \psi_{1,2}(x,t)=\frac{1}{\sqrt{\rho(t)}}\,e^{-i\omega_I(x,t)}\phi_{1,2}(y,t)
\equiv \Omega (x,t) \phi_{1,2}(y,t)
\end{equation}
this Schr\"odinger equation becomes quasi-stationary \cite{Bluman96} 
\begin{equation}
  i\rho^{2}(t)\partial_{t} \phi_{1,2}(y,t)=
  \left[-\partial_{y}^{2} + K'^{2}(y) \pm K''(y)\right]\phi_{1,2}(y,t)\;,
\label{SEiny}
\end{equation}
which is obviously separable in $y$ and $t$. Hence, the solutions of the
original Schr\"odinger equations have the general form
$\psi(x,t)=\Omega(y,t)Y(y)T(t)$ which 
is known as the $R$-separation of variables \cite{Miller77}. In other words,
for any pair of  
Schr\"odinger operators ${\cal S}[V_{1,2}]$, which admits a first-order intertwining
relation (\ref{inS}) there exists a transformation (\ref{trafo}) to some new
coordinate in which the potentials become stationary (see also \cite{Finkel98}). 
Notice that the transformation
associated with the special case $\rho(t)=1$ and $\mu(t)=vt$ with constant
velocity $v$ corresponds to the Galileo transformation. See, for example, the
textbook \cite{Galindo90}.

 This $R$-separation of variables is certainly
related to the existence of a symmetry operator. 
First,
one can directly verify the adjoint intertwining relation for  real potentials,
\begin{equation}
  q_t^ -{\cal S}[V_1]={\cal S}[V_2]q_t^ -
\label{inSadjoint}
\end{equation}
where
\begin{equation}
q_t^ -\equiv(q_t^ +)^\dagger=
  -\rho(t)\partial_x+K'\Bigl(x/\rho(t)+\mu(t)\Bigr)+
  \frac{i}{2}\left(\dot{\rho}(t)x-\rho^2(t)\dot{\mu}(t)\right).
\end{equation} 

Then from (\ref{inS}), \gl{susytime} and (\ref{inSadjoint}) we obtain the closure of
the SUSY algebra,
\begin{equation}
\left\{Q_t, Q^{\dagger}_t\right\} = {\cal R}_t,\quad 
  \Bigl[\hat{\cal S}_t, {\cal R}_t\Bigr]=0\;,\quad Q^{\dagger}_t =
  (Q_t)^\dagger\, ,
\end{equation}
where the symmetry operator ${\cal R}_t$ has the following components
\ba
\fl R_t^\pm = q_t^\pm q_t^\mp =  -\rho^2(t)\partial^2_x + 
\frac{i}{2}\left(\rho \dot{\rho}(t)\{x,\partial_x\}- 
2\rho^3(t)\dot{\mu}(t)\partial_x\right)\nonumber\\ +
\left[
  K'\Bigl(x/\rho(t)+\mu(t)\Bigr)\right]^2
 \pm K''\Bigl(x/\rho(t)+\mu(t)\Bigr) + \frac{1}{4}
\left(\dot{\rho}(t)x-\rho^2(t)\dot{\mu}(t)\right)^2\nonumber\\
\lo= \exp\{-i\omega_I (x,t)\} \left[-\partial_{y}^{2} +
  K'^{2}(y) \pm K''(y)\right] \exp\{i\omega_I (x,t)\}.
\ea
Thus the quasi-stationary Hamiltonians in Eq.~\gl{SEiny} are just
unitary
equivalent to the symmetry operators $R^\pm_t$. It means that the
supersymmetry entails the separation of variables because it provides a
new symmetry. As a consequence the quantum
dynamics splits in orbits with a given eigenvalue of the symmetry operator. 
 %%%%%%%%%%%%%%%%%%%%%%%%%%%%%%%%%%%%%%%%%%%%%%%%%%%%%%%%%%%%%%%%%%%%%%%%%%%%%%
%%%%%%%%%%%%%%%%%%%%%%%%%%%%%%%%%%%%%%%%%%%%%%%%%%%%%%%%%%%%%%%%%%%%%%%%%%%%%%
\section{Second-order intertwining for stationary potentials:
symmetry operators and spectrum generating algebra}

\hspace*{3ex} Now we will report on the intertwining of a
pair of Schr\"odinger operators ${\cal S}[V_1]$ and ${\cal S}[V_2]$ by second-order
(intertwining) operators of the form: 
\begin{equation}
  q_t^{+}(x,t) = G(x,t)\partial_{x}^{2} - 2F(x,t)\partial_{x} + B(x,t)\;.
\label{old18}
\end{equation}
We will explore the connection of the time-dependent SUSY charges with 
appearance of the spectrum generating (oscillator like) algebras for  
the corresponding Hamiltonians.

As in the first-order case it can be shown \cite{tdi} 
that the inclusion of an additional
term being of first order in $ \partial_{t} $ leads to the trivial situations
where the difference $V_1-V_2$ depends on the time $t$ only. Furthermore, from the intertwining relation
(\ref{inS}) with above $q_t^ +$ one can conclude that the function $G$ may not
depend on $x$ and similarly to the discussion in the previous section it is even
possible to exclude a phase. In other words, without loss of generality 
 $G(x,t)\equiv g(t)$ and consider from now on an intertwining operator of the
form 
\begin{equation}
  q_t^{+}(x,t) = g(t)\partial_{x}^{2} - 2F(x,t)\partial_{x} + B(x,t)\;.
\label{18}
\end{equation}

In \cite{tdi} particular solutions of the intertwining
relation (\ref{inS})  were constructed with $q_t^ {+}$ as given above . 
In this section we shall analyze 
 the solutions
of the intertwining relation (\ref{inS}) for the case where both potentials
$ V_1$ and $V_2$ are stationary, {\it i.e.} do not depend on $t$.
 
One class of such solutions is known from 
\cite{acdi95}. Assuming a supercharge $ q_t^{+} $ 
with real coefficient functions independent on $t$, one finds that
the corresponding solutions of (\ref{inS})  coincide with those of
the stationary intertwining relations $(-\partial^2_x + V_1(x))q^{+}(x) =
q^{+}(x)(-\partial^2_x + V_2(x))$ from \cite{acdi95}.
 
Here we are interested in more general solutions of (\ref{inS}) when
operators $ q_t^{+}$ depend  on $t$,
\begin{equation}
  (i\partial_t -h^+)q_t^+(x,t)=q_t^+(x,t) (i\partial_t -h^-)\;,
\label{inH}
\end{equation}
with standard stationary Hamiltonians $h^\pm=-\partial_x^2+V_{1,2}(x)$ but
explicitly time-dependent intertwining operators.

Let us employ the suitable ansatz with simple $ t$-dependence in (\ref{18}),
\begin{equation}
  q_t^ {+}(x,t) = M^{+}(x) + A(t)a^{+}(x)\;, 
  \label{20}
\end{equation}
where 
\begin{equation}
  M^{+}(x)\equiv \partial_{x}^{2} - 2f(x)\partial_{x} + b(x)\;,\qquad
  a^{+}(x)\equiv \partial_{x} + W(x)\;. \label{20a}
\end{equation}
Here all functions besides $A$ are considered to be real.  We
  also assume $A\not\equiv 0$.  
With this ansatz the intertwining relation (\ref{inH}) can be shown \cite{tdi} 
to yield

\begin{equation}
  \begin{array}{l}
i\dot A=2\,\tilde m + 2\,m\, A\;,\\[2mm]
h^+M^{+} - M^{+}h^- = 2\,\tilde m\, a^{+}\;,\\[2mm]
h^+a^{+} - a^{+}h^- = 2\, m\, a^{+}\;,
  \end{array}\label{25}
\end{equation}
with real constants $ \tilde m$ and $m$. 

We find it interesting to focus on the case $m\neq 0$ to explore
certain spectrum generating algebras.
 The first equation in (\ref{25}) immediately leads to
\be
A(t) = m_0e^{-2imt} - \tilde m/m \label{time}
\ee 
with a real $m_0$, and 
\begin{equation}
  q_t^ {+}(x,t) = \partial_{x}^{2} - \biggl(2f(x) + \frac{\tilde m}{m}
\biggr)\partial_{x} + b(x) - \frac{\tilde m}{m}\,W(x) + m_0e^{-2imt}a^{+}(x)\;.
\end{equation}
It is obvious that without loss of generality we may set $\tilde m = 0$
as a non-vanishing $\tilde{m}$ may always be absorbed via a proper
redefinition of $f$ and $b$, {\it i.e.} of the operator $M$. 

As a consequence, the second relation in
(\ref{25}) leads to a second-order intertwining between $h^+$ and $h^-$.  
This has already been considered in \cite{acdi95} and it was found that 
the potentials $V_1$, $V_2$ and the function $b$ can be expressed in terms of
$f$ and two arbitrary real constants $a$ and $d$:
\begin{equation}
  \begin{array}{l}
\displaystyle
V_{1,2}(x) = \mp 2 f'(x) + f^{2}(x) + \frac{f''(x)}{2f(x)} 
         - \frac{f'^{\, 2}(x)}{4f^{2}(x)}- \frac{d}{4f^{2}(x)} - a\;,\\[2mm]
\displaystyle
b(x) = -f'(x) + f^{2}(x) - \frac{f''(x)}{2f(x)} 
       + \frac{f'^{\, 2}(x)}{4f^{2}(x)} + \frac{d}{4f^{2}(x)}\; .
  \end{array}\label{27}
\end{equation}
The corresponding second-order SUSY algebra generated by the
supercharge $M$ is similar to \gl{secor},
\be
\{M,\, M^\dagger\} = (H + a)^2 + d \equiv {\cal P}_2 (H) .
\la{secord}
\ee
Note that cases with $d<0$
are reducible ones. 

One may find some similarities between the present intertwining
algebra \gl{25} and the extended SUSY relations discussed in Section
5. But we emphasize that now for $m\not= 0$ the last relation in
\gl{25} does not generate a second SUSY. Rather it creates the
equivalence of relatively
shifted spectra of two Hamiltonians $h^+$ and $h^-$ which is typical for
spectrum generating algebras. Specifically
\be
a^+ a^- = h^+ - m + c;\quad a^- a^+ = h^- + m +c, \label{shifted}
\ee 
where $c$ is a real constant.
Therefore the
reflectionless potentials found in Section 5 are produced only in the limit
of $m = 0$. For this reason we use here the notations for relevant
operators different from those ones in Section 5.

The genuine spectrum generating algebra for stationary Hamiltonians
$h^\pm$ can be derived from Eq.~\gl{25}
\begin{equation}
  \begin{array}{l}
[h^+, G_+] = - 2m G_+ \;,\quad G_+ \equiv
M^{+}a^{-}\, ,\\[2mm]
[h^+, G^\dagger_+] =  2m  G^\dagger_+\;,\quad G^\dagger_+ \equiv a^{+}
M^- \, , \\[2mm]
[h^-,  G^\dagger_-] = 2m  G^\dagger_-  \;,\quad G^\dagger_- \equiv
M^{-}a^{+} \, , \\[2mm]
[h^-, G_-] = - 2m G_- \;, \quad G_- \equiv a^{-}
M^+ \, ,
  \end{array}\label{sga}
\end{equation}
where $a^-=(a^+)^\dagger$ and $M^-=(M^+)^\dagger$.
The closure of this spectrum generating algebra  is a polynomial
deformation of Heisenberg algebra \cite{ferhus2},
\be
[G^\dagger_\pm, G_\pm] = F^\pm (h^\pm).
\ee
The explicit form of the polynomials $F^\pm(x)$ can be obtained with
the help of Eqs.~\gl{25} and \gl{shifted} for $\tilde m = 0$. For
instance,
\ba
G^\dagger_+ G_+ =  (h^+ -m +c)\, {\cal P}_2 (h^+ - 2m) ;\nonumber\\
G_+ G^\dagger_+ =  (h^+ +m +c)\, {\cal P}_2 (h^+) , 
\ea
where the notations from Eqs.~\gl{secord} and \gl{shifted} are
employed.
The polynomials $F^\pm(x)$ 
turn out to be different for the isospectral partners
$h^\pm$,
\ba
\fl F^+(h^+) = - 6m (h^+)^2 + 4m(2m - 2a -c) h^+ -2m[a^2 + d + 
2(a-m)(c-m)];\nonumber\\
\fl F^-(h^-) = - 6m (h^-)^2 - 4m(2m + 2a +c) h^- -2m[a^2 + d + 
2(a+m)(c+m)] .
\ea
Hence the two spectrum generating algebras are, in general, different
that is essentially due to the shift in intertwining relations 
\gl{shifted}. 
There is a formal discrete symmetry between their constants and
Hamiltonians $h^+, a, c \Longrightarrow - h^-, - a, -c$.

 The intertwining
relation (\ref{inH}) and its adjoint give rise to the symmetry operators
$q_t^ {+}q_t^ {-}$ and $q_t^ {-} q_t^ {+}$ for $(i\partial_{t} - h^+)$ and
$(i\partial_{t} - h^-)$, respectively. 
Using Eqs.~\gl{time},\gl{secord},\gl{shifted} and after elimination of
polynomials 
of the Hamiltonians $h^\pm$ 
these symmetry
operators may be reduced to the 
form 
\begin{equation}
  \begin{array}{l}
R^+(x,t) =m_0\left[ e^{2imt} G_+ + e^{-2imt} G^\dagger_+\right]\;,\\[2mm]
R^-(x,t) = m_0 \left[e^{2imt} G_- + e^{-2imt} G^\dagger_-\right]\;.
  \end{array}\label{30}
\end{equation}
 As our potentials
do not depend on time the operators $R^\pm(x,t +\Delta)$ with a time
shift $\Delta$  are also symmetry operators for the same 
Schr\"odinger equation,
\be
\left[R^\pm_{(t +\Delta)}, S_t\right] = 0.
\ee
In particular, time derivatives $\dot{R}^\pm(x,t)$
of hermitian symmetry operators $R^\pm(x,t)$ form an independent set of
hermitian symmetry operators which do not commute between themselves.
Similar
results have also been obtained in \cite{nikitin}
using a different approach. We see that the non-stationary SUSY 
delivers the symmetry operators which encode the entire set of
spectrum generating algebras
\be
e^{2imt} G_\pm = \frac{1}{2m_0} R^\pm(x,t) -\frac{i}{4m m_0} \dot R^\pm(x,t). 
\ee

The natural question  concerns 
the reducibility
of the second-order intertwining operator (\ref{18}) to a pair of consecutive
first-order operators. 
Progress in classification of possible irreducible transformations has
been made in \cite{tdi} though  more work must be done toward the
full classification.

\section{Conclusions and perspectives}
\hspace*{3ex} The purpose of this short review has been 
two-fold: to elucidate the
recent progress in Nonlinear SUSY realization
 for a broad community of spectral designers and to
draw reader's attention to a variety of SUSY extensions which yield
new QES potential systems and illuminate 
some old ones.
With the experience from the previous sections 
the general SUSY QM 
can be thought of as governed by the extended nonlinear SUSY algebra
with ${\cal N}$ pairs of nilpotent supercharges $Q_j,\, Q_j^\dagger $ 
and a number of hermitian hidden-symmetry differential
operators 
$R_\alpha = R_\alpha^\dagger,\quad [R_\alpha, R_\beta] = 0;\quad 0\leq
\alpha,\beta \leq M$. Such a SUSY algebra takes the modified form,
\ba
\left[R_\alpha, Q_k\right]= \left[R_\alpha, Q^{\dagger}_k\right] = 0;\quad 
\left\{Q_j, Q_k\right\} = \left\{Q^{\dagger}_j, Q^{\dagger}_k\right\} = 0;\nonumber\\
\left\{Q_j, Q^{\dagger}_k\right\} = {\cal P} (R_\alpha); \label{genclos}
\ea
We notice that, first, the Super-Hamiltonian itself is included into the set of
symmetry operators, say for $\alpha = 0$, $R_0 \equiv H$ and, second,
not all the symmetry operators are necessarily  present in the
algebraic closure \gl{genclos} (see Sections 7,8).
 
On the other hand, 
there remains a plenty of open questions and challenges to be solved. 
\begin{itemize}
\item In the first half of this paper we have given a systematic
analysis of reducibility vs. irreducibility of 
type I, II, III in the one-dimensional QM. 
However the higher-order irreducibility needs more efforts to prove
the exhaustive completeness of the classification in Section 3. 
\item For SUSY extensions it may be of interest to find pairs 
of (quasi)isospectral potentials admitting hidden symmetries which are
related by a type-II irreducible Darboux transformation.
\item The irreducibility
classification for non-stationary potentials as well as the existence
of extended
SUSY   is very welcome to be
investigated and new applications to be found, in particular, to
explore spectrum generating algebras (see Section 8).
\item The similarity of the Schr\"odinger equation to the
  Fokker-Planck one allows \cite{junker,tdi,miof} to find
 the SUSY scheme for generating new solutions of the latter equation. One
  can 
be tempted to develop a more exhaustive analysis of how to produce
  SUSY clones using the ideas of the conventional nonlinear SUSY
outlined
here. But attention should be focused on the 
non-hermiticity of the Fokker-Planck operator and on the fact
that the equivalent of the wave function is a positive and properly 
normalized probability function.
\item  Matrix (coupled channel) systems represent a rich and not fully
scanned field of extended SUSY systems with hidden
symmetries.
While certain interesting matrix potentials have been explored
\cite{Amado:1988jn},
\cite{matrix1}-\cite{matrix4}
it is clear that in this case  the way to a comprehensive understanding of
irreducible
building blocks for spectral design is still long. 
\item The polynomial SUSY in two dimensions has already
  brought  a number of examples of new type of irreducible SUSY with 
hidden symmetries of higher-order in derivatives 
\cite{ain95,aintmf,ioffe:jpa}.  One may expect a variety of new types of
irreducible SUSY for third-order (and higher-order) supercharges as
  well as new discoveries in three dimensions.
\item Complex potentials (\cite{complex0}--\cite{complex4}) seem to offer
  less problematic generalizations of QM with hermitian Hamiltonians as
  compared to matrix and multi-dimensional QM. 
Therefore many of tools and results of one-dimensional SUSY QM are
  expected to be applicable when a potential is complex \cite{ansok}.
However, as it was recently remarked in \cite{mostafa} there exist non-hermitian
  Hamiltonians which are not diagonalizable but at best can be reduced
  to a Jordan form. For the latter ones special care must be taken to
derive the isospectrality and to build SUSY ladders.
\end{itemize}
\ack
We are very grateful to the Organizers of the Conference ``Progress in
Supersymmetric Quantum Mechanics-2003'' for hospitality in Valladolid 
and especially
to L.M.Nieto and J.Negro for their patient and friendly 
encouragement to complete this paper. We thank also A.V.Sokolov for
numerous
discussions of irreducibility problems. A.A. has been supported by the
Russian Fond for Basic Research (Grant No.02-01-00499).
%%%%%%%%%%%%%%%%%%%%%%%%%%%%%%%%%%%%%%%%%%%%%%%%%%%%%%%%%%%%%%%%%%%%%%%%%%%%
%%%%%%%%%%%%%%%%%%%%%%%%%%%%%%%%%%%%%%%%%%%%%%%%%%%%%%%%%%%%%%%%%%%%%%%%%%%%%%%

\end{document}